# Fourier optical processing enables new capabilities in diamond magnetic imaging

Mikael P. Backlund[1, 2], Pauli Kehayias[1, 2], and Ronald L. Walsworth[1, 2*]

1. Harvard-Smithsonian Center for Astrophysics, Cambridge, Massachusetts 02138, USA
2. Department of Physics, Harvard University, Cambridge, Massachusetts 02138, USA
*rwalsworth@cfa.harvard.edu

**Diamond-based magnetic field sensors have attracted great interest in recent years[1, 2]. In particular, wide-field magnetic imaging using nitrogen-vacancy (NV) centers in diamond has been previously demonstrated in condensed matter[3], biological[4, 5], and paleomagnetic[6, 7] applications. Vector magnetic imaging with NV ensembles typically requires an applied field (>10 G) to separate the contributions from four crystallographic orientations, hindering studies of magnetic samples that require measurement in low or independently specified bias fields. Here we decompose the NV ensemble magnetic resonance spectrum without such a bias field by modulating the collected light at the microscope's Fourier plane[8]. In addition to enabling vector magnetic imaging at arbitrarily low fields, our method can be used to extend the dynamic range at a given bias field. As demonstrated here, optically-detected diamond magnetometry stands to benefit from Fourier optical approaches, which have already found widespread utility in other branches of photonics.**

NV centers are $C_{3v}$-symmetric color centers of the diamond lattice formed by substitution of a nitrogen atom and a vacancy at neighboring lattice sites (Fig. 1a-c). The negatively-charged NV center has an electronic spin-triplet ground state with a 2.87-GHz zero-field splitting between the $m_s = 0$ and $m_s = \pm 1$ magnetic sublevels (Fig. 1d). A magnetic field **B** further splits these sublevels by the Zeeman effect. Illuminating an NV center with 532-nm laser light optically pumps it to the $m_s = 0$ sublevel, which luminesces more brightly than the $m_s = \pm 1$ sublevels. A resonant microwave field repopulates the $m_s = \pm 1$ sublevels and reduces the photoluminescence (PL) intensity. Measuring PL as a function of microwave frequency yields an optically-detected magnetic resonance (ODMR) spectrum, from which one can calculate **B**.

In an NV ensemble there are typically equal populations of the four possible NV orientations within the diamond crystal, with spin quantization axes parallel to $\{\hat{u}_1, \hat{u}_2, \hat{u}_3, \hat{u}_4\}$ (Fig. 1a, b). Different projections of **B** result in a total of eight ODMR spectral lines for the four orientations (Fig. 2b). To resolve these eight resonances one must typically apply a sufficiently strong bias field **B**[(bias)]. However, a strong applied field is undesirable for some applications, including paleomagnetic studies of geological and meteorite samples in which induced signal from paramagnetic and low-coercivity grains may overwhelm the ferromagnetic signal of interest[6, 9]. In other applications it may be necessary to reserve the applied field for another function, e.g. for controlling magnetic nanoparticles within a cell[10] or tuning the properties of a magnetic material[11, 12].

Here we present a practical, all-optical method to resolve the NV ensemble ODMR spectrum using Fourier plane processing, exploiting the selection rules of the $^3E \rightarrow {}^3A_2$ optical transition (Fig. 1d)[13-15]. In the paraxial regime, optical polarization alone cannot distinguish emission from orientation pairs $\hat{u}_j$ and $\hat{u}_{j+2}$. To gain deeper insight we simulated NV PL for the four orientations by adapting previous work modeling electric dipole emission near interfaces[16-21]. We model emission of a given orientation as arising from two mutually-incoherent, degenerate electric dipole emitters oriented perpendicular to the NV axis (Fig. 1c). Our simulations predict a characteristic distribution of intensity and contrast for each orientation at the microscope's Fourier plane (Fig. 1f; intuition given in Fig. S1). For experimental confirmation we employed the setup sketched in Fig. 1e, using a Bertrand lens to relay the Fourier plane onto the camera and a linear polarizer inserted into the collection path to pass either *x*- or *y*-polarized PL. We visualized the Fourier-plane distribution of ODMR contrast due to each NV orientation by integrating the corresponding resonances resolved by a sufficiently strong $\mathbf{B}^{(\text{bias})}$. Figure 1g shows these results, indicating excellent agreement between simulation and experiment. We next sought to leverage this effect to decompose the ODMR for general $\mathbf{B}^{(\text{bias})}$.

Fourier ODMR decomposition is achieved by making four sequential measurements (Fig. 2a). The NV ensemble ODMR spectrum recorded with an *x*-oriented linear polarizer and the left half of the pupil blocked is $\tilde{c}_1(f)$; $\{\tilde{c}_2(f), \tilde{c}_3(f), \tilde{c}_4(f)\}$ are defined similarly. Together these four measurements comprise $\tilde{\mathbf{C}} \in \mathbb{R}^{4 \times N_f}$, where $N_f$ is the number of microwave frequencies sampled. Note that $\langle \tilde{c}_i(f) \rangle = \frac{1}{4} \sum_i \tilde{c}_i(f)$ represents a conventional ODMR spectral measurement without filtering (Fig. 2b, c). In Fig. 2 we present two data sets, one in which $\mathbf{B}^{(\text{bias})}$ resolves the resonances of each orientation (Fig. 2b, d, f), and one in which it does not (Fig. 2c, e, g). We define $\Delta_B$:

$$\Delta_B = \max \left| \mathbf{B}^{(\text{bias})} \cdot \mathbf{u_j} \right| - \min \left| \mathbf{B}^{(\text{bias})} \cdot \mathbf{u_j} \right|, \tag{1}$$

enumerating the range in projections of the bias field onto the NV orientations, and the degree of overlap of the resonances. For data in Fig. 2b, d, f $\Delta_B$ = 19.52 G, while $\Delta_B$ = 0.42 G for data in

Fig. 2c, e, g. The latter value is not a fundamental minimum for $\Delta_B$ and was only chosen qualitatively during the experiment to overlap the ODMR peaks. Throughout our studies we applied a bias field such that $\min \left| \mathbf{B}^{(\mathbf{bias})} \cdot \mathbf{u_j} \right| > 1.4$ G in order to simplify our analysis and focus on the technological advancement at hand. At lower $\min \left| \mathbf{B}^{(\mathbf{bias})} \cdot \mathbf{u_j} \right|$, ODMR spectra become complicated by level crossings as the Zeeman splitting approaches those due to the $^{14}$N hyperfine interaction[22]. This issue can be addressed by $^{15}$N enrichment and controlled circular microwave polarization[23-25], which is compatible with our all-optical technique. We reserve the technical work of combining these methods for future studies.

We first demonstrated our method by directly imaging the Fourier plane, taking sequential ODMR measurements of different polarizations, and integrating halves of the pupil plane in post-processing. Individual $\tilde{c}_i(f)$ are shown in Fig. 2d, e for high and low $\Delta_B$, respectively. From simulation, we expect that for each $\tilde{c}_i$, the resonances should have relative weights $w_1 = 0.226$, $w_2 = 0.363$, $w_3 = 0.049$, and $w_4 = w_2$. Experimentally measured weights were $w_1 = 0.220(4)$, $w_2 = 0.360(7)$, and $w_3 = 0.060(2)$. We define $\mathbf{C} \in \mathbb{R}^{4 \times N_f}$, containing the underlying ODMR spectrum $c_j(f)$ for each NV orientation, yielding

$$\tilde{\mathbf{C}} = \mathbf{W}\mathbf{C}, \qquad (2)$$

where $\mathbf{W}$ is the circulant matrix[26] formed by permuting $[w_1, w_2, w_3, w_2]$. The underlying isolated ODMR spectrum of each individual orientation can be estimated:

$$\hat{\mathbf{C}} = \mathbf{W}^{-1}\tilde{\mathbf{C}}. \qquad (3)$$

Thus, while the high-$\Delta_B$ demonstration in Fig. 2f proves the capability of isolating ODMR features of individual NV orientations, Fig. 2g demonstrates this capability in the more useful low-$\Delta_B$ regime.

To demonstrate wide-field imaging with Fourier plane modulation, we removed the Bertrand lens and added the lens L2 (Fig. 1e), forming a 4$f$ optical processing unit[8] between which we placed a knife-edge beam block to alternately obscure halves of the pupil. While the beam block modifies the microscope's point-spread function (PSF) (Fig. S2), an image is nonetheless relayed to the camera and the measurement procedure yields the ODMR spectrum $\tilde{c}_i(x_k, y_k, f)$ in each pixel $k$. After pixelwise transformation via Eq. 3, each $c_j(x_k, y_k, f)$ spectrum is fit and the vector magnetic field is reconstructed across the image. We applied this procedure to image the field from a ferromagnetic bead placed directly on the surface of a diamond containing a 3.8-μm NV layer (Fig. 3; additional beads in Fig. S3). To establish a ground truth, we first resolved the resonances of each orientation (Fig. 3a) at high $\Delta_B$ and inferred the sample field (Fig. 3b-d) in the conventional way[9]. Next we reduced $\Delta_B$ to overlap the resonances (Fig. 3e). Using our method we decomposed the unresolved spectrum and inferred the vector magnetic field image (Fig. 3f-i). Comparable fits to magnetic dipole sources are shown in the insets of Fig. 3b-d, f-h. This demonstration shows that the Fourier optical decomposition method can be used to accurately reconstruct vector magnetic images at fields for which overlapping resonances would otherwise lead to ambiguity. We emphasize that the utility of our method is in resolving such ambiguities—relative to NV vector magnetic imaging at large bias field, our method yields reduced sensitivity (Fig. S4 and supplementary text), in part due to the sequential nature of the present measurement. In future implementations, some of this difference can be recovered by parallelization.

As a final demonstration, we imaged a 30-μm thin section of the Allende CV3 chondrite, a widely studied meteorite thought to be magnetized by a possible dynamo of its parent planetesimal[9, 27]. For this proof-of-principle we enhanced the magnetic signal by applying a large isothermal remanent magnetization (IRM) field to the sample before imaging. An external bias field with $\Delta_B$ = 8.42 G was applied in order to just resolve the NV ODMR peaks in the absence of the sample's field (Fig. S5). The ODMR peaks shift and overlap considerably in some regions, collapsing into two or four overlapped peaks and leading to ambiguities in the field components when treated without decomposition. While sample field reconstruction fails with the conventional technique (Fig. S5), our Fourier decomposition technique allows us to determine vector magnetic images (Fig.4a-c), revealing a pattern resembling two strong dipolar features atop a slowly varying background field (Fig. S6). Figure 4d, e shows a brightfield image mapping the positions of these two dipolar sources to the edge of a chondrule. Figure 4f, g gives further insight, displaying the ODMR spectra with and without decomposition along a slice of the image. Our method effectively increases the dynamic range in this measurement by >3x (Fig. S5), without the need to increase $B^{(bias)}$. This capability is useful for samples that can withstand modest $B^{(bias)}$ but for which exceedingly large $B^{(bias)}$ remain undesirable.

In conclusion, we demonstrated a Fourier optical decomposition method to realize NV ensemble vector magnetic imaging without need of a large bias magnetic field to resolve the contributions from each NV orientation. This method enables vector magnetic imaging in applications where a large bias field may induce unwanted magnetization the sample, or must be reserved for other experimental functions. It can also be used to extend dynamic range at any $B^{(bias)}$. The present work represents one of the first applications of Fourier optical processing to NV imaging. This class of techniques has found broad utility in other realms of optical imaging[8, 21, 28, 29], and has great potential for future NV applications.

---

We recently became aware of an alternative approach that simultaneously exploits optical and microwave absorption selection rules for NV ensemble vector magnetometry[30]. We note that at very low applied fields (< 100 mG) microwave polarization control must be reserved to distinguish transitions to $m_s = \pm 1$, and so our all-optical method is compatible with imaging small sample fields under such conditions.

**Acknowledgements**

This material is based upon work supported by, or in part by, the U. S. Army Research Laboratory and the U. S. Army Research Office under contract/grant number W911NF1510548, as well as by the NSF EPMD, PoLS, and INSPIRE programs.  P. K. acknowledges support from the Intelligence Community Postdoctoral Research Fellowship Program. We thank Benjamin Weiss for the Allende sample and for use of his reflection microscope, Eduardo Lima for help with brightfield reflection microscopy, and Roger Fu for performing IRM on the Allende sample and for helpful discussions. We also thank David Glenn, Mark Ku, John Barry, Jennifer Schloss, and Matthew Turner for helpful discussions.

**Author contributions**

M.P.B. did calculations and analysis. M.P.B. and P.K. did the experiments. M.P.B., P.K., and R.L.W. wrote the paper.

**Competing financial interests**

The authors declare no competing financial interests.

**Figures**

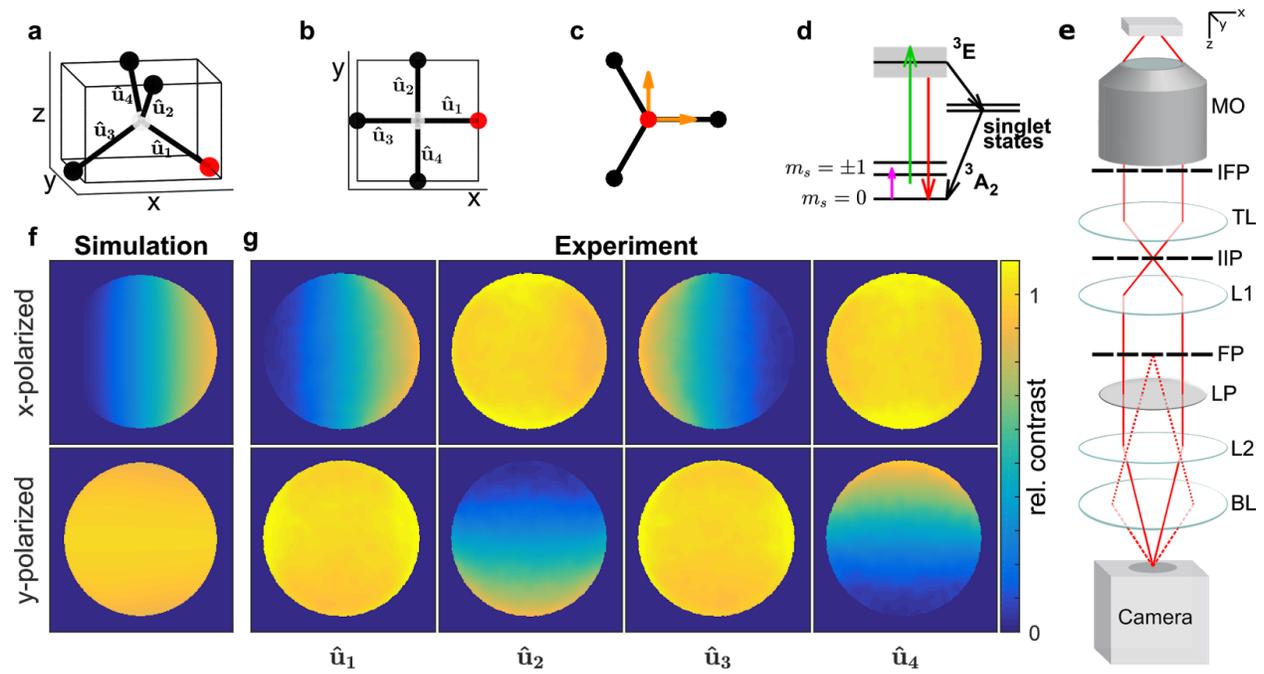

**Figure 1 | Setup and Fourier plane patterns of NV ensemble photoluminescence. a**, Sketch of NV center pointing along $\hat{u}_1$, including 3 carbon atoms (black spheres), nitrogen atom (red), and vacancy (gray). Lines parallel to each NV orientation class $\hat{u}_j$ are labeled. Facets of the illustrated rectangular prism coincide with those of the diamond samples used in our experiments and are perpendicular to $[110]$, $[1\bar{1}0]$, and $[001]$. **b**, $xy$ perspective of lattice shown in **a**. **c**, View of NV along its axis, with transition electric dipole moments (orange). **d**, Simplified NV energy level diagram (level spacings not to scale). Electronic $^3A_2$ ground state consists of $m_s = 0$ and $m_s$

= ±1 magnetic sublevels split by $D$ = 2.87 GHz. The Zeeman effect further splits $m_s$ = ±1 in response to a magnetic field. Applied microwaves (magenta) facilitate transitions from $m_s$ = 0 to $m_s$ = ±1. A 532-nm laser is applied to drive optical transitions to the $^3E$ manifold. Red photoluminescence (PL) is collected upon radiative relaxation to the ground state. Nonradiative relaxation through the singlet state channel is responsible for the spin-state dependent PL contrast and initialization into $m_s$ = 0. **e**, Schematic of optical setup, tracing collected PL from the diamond chip to the camera: microscope objective (MO), intermediate Fourier plane (IFP), tube lens (TL), intermediate image plane (IIP), 4*f* lens (L1), Fourier plane (FP), linear polarizer (LP), and either 4*f* lens (L2) or Bertrand lens (BL) depending on whether the measurement calls for imaging real of Fourier space. **f**, Simulation of PL contrast distribution in the Fourier plane for an NV oriented along $\hat{u}_1$ with an NA = 1.49 oil objective. **g**, Experimentally measured NV ensemble PL contrast distributions in the Fourier plane due to each NV orientation.

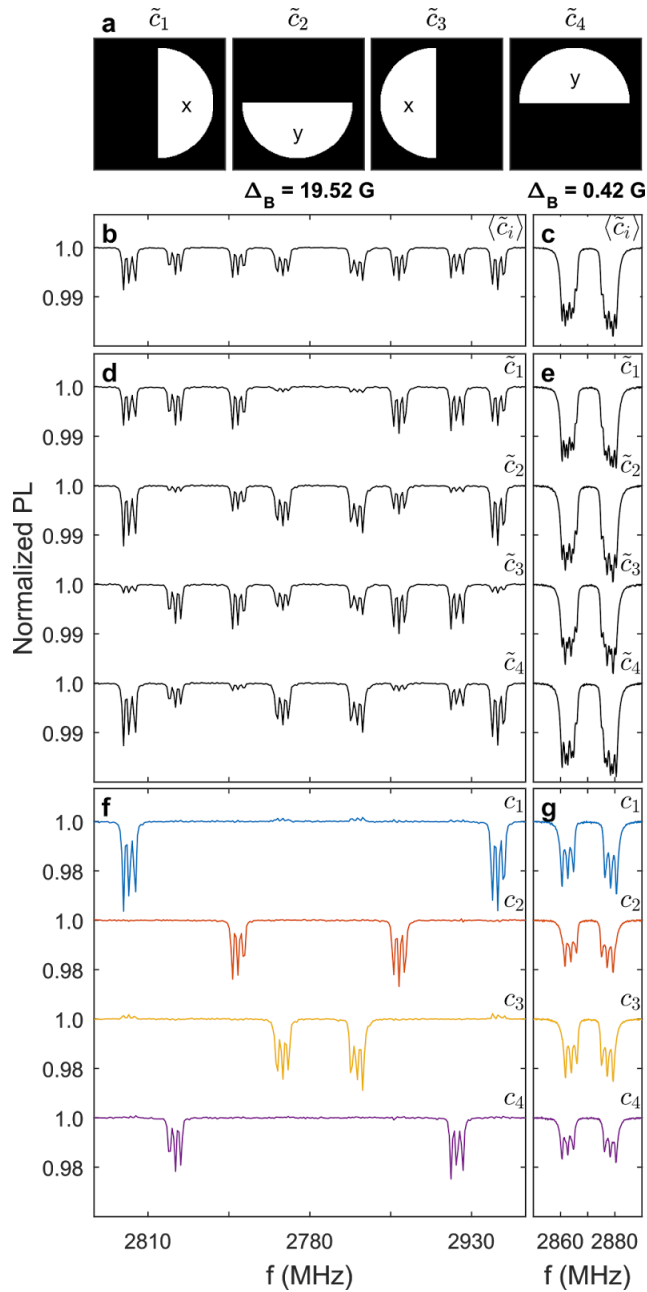

**Figure 2 | Fourier decomposition of optically-detected magnetic resonance spectrum. a**, Schematic depicting the passed polarization as well as the integrated/discarded portions of the pupil for measurement of each spectrum $\tilde{c}_i(f)$. **b**, Conventional NV ensemble optically-detected magnetic resonance (ODMR) spectrum realized by computing the average of four measurements

$\langle \tilde{c}_i(f) \rangle$ at high $\Delta_B$ (= 19.52 G) such that eight resonances are clearly resolved. Each resonance appears as a triplet due to ~2.16 MHz splitting from the $^{14}$N hyperfine interaction. **c**, Same as **b** but at low $\Delta_B$ (= 0.42 G) such that resonances from different NV orientations are not resolved. **d**, ODMR measured under each of the four conditions sketched in **a** at $\Delta_B$ = 19.52 G. **e**, Same as **d**, but instead at $\Delta_B$ = 0.42 G. **f**, Resulting ODMR spectra after Fourier decomposition at $\Delta_B$ = 19.52 G. Average crosstalk error is ~1%, as determined by comparing absolute value of the integral under each would-be nulled resonance to the non-nulled resonance. **g**, Same as **f**, but instead at $\Delta_B$ = 0.42 G. The highly overlapping ODMR peaks obscure the hyperfine splittings in **c** and **e**, whereas **g** shows that they are clearly revealed by the transformation. Different spectra within the same panel in **d-g** are offset for clarity.

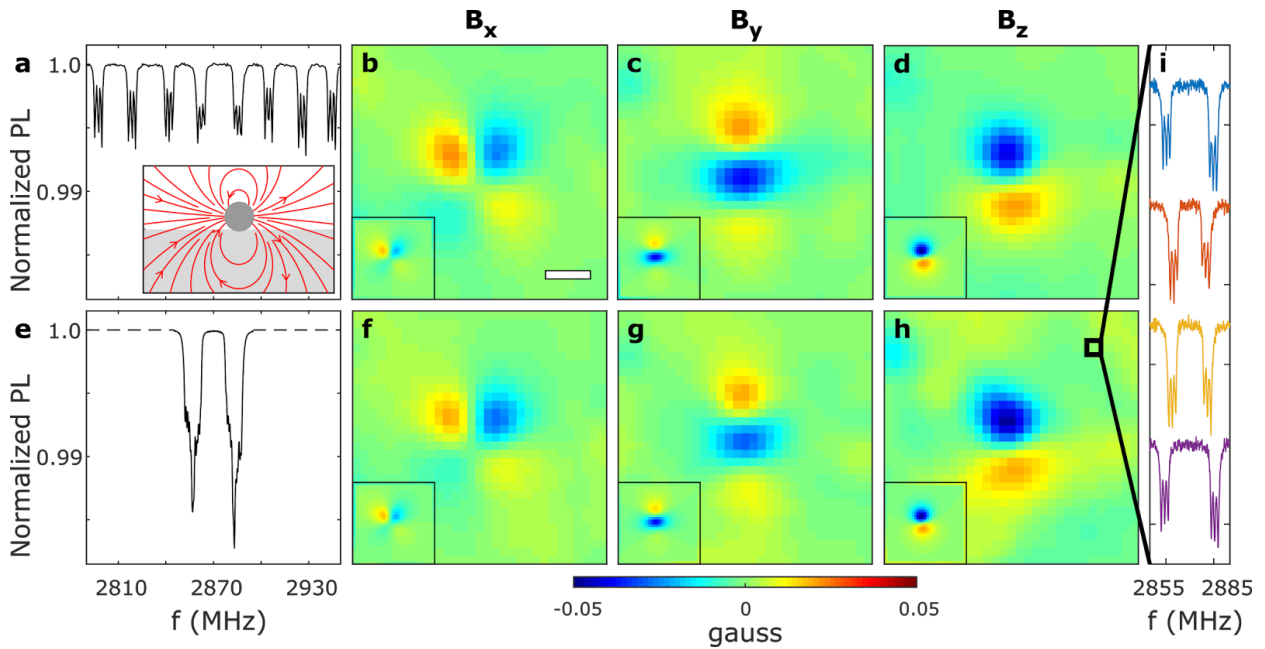

**Figure 3 | Magnetic bead imaging with Fourier optical decomposition. a**, Spatially-averaged NV ensemble ODMR spectrum at $\Delta_B$ = 22.16 G such that resonances are well resolved. Inset:

cartoon of magnetic bead on diamond surface with magnetic field lines (red). **b-d**, images of $x$, $y$, and $z$ components of magnetic field due to a magnetic bead, determined at $\Delta_B = 22.16$ G without optical decomposition. Insets show calculated field components from fit to magnetic dipole source, with fit $x$ position -0.2(1) µm, $y$ position -0.3(1) µm, standoff distance = 8.9(1) µm, magnetic dipole moment = 29(1)×10$^{-15}$ J/T, azimuthal orientation = 82(1)°, and polar orientation = 110(1)°. Fit parameter errors are average 95% confidence intervals determined using MATLAB function `nparci`. Scale bar: 10 µm. **e**, Spatially-averaged NV ensemble ODMR spectrum from measurement at $\Delta_B = 1.99$ G such that resonances are not resolved. **f-h**, Images of $x$, $y$, and $z$ components of stray magnetic field due to same magnetic bead as in **b-d**, but determined at $\Delta_B = 1.99$ G with Fourier optical decomposition. Insets show calculated field components from fit to magnetic dipole source, with fit $x$ position -0.6(1) µm, $y$ position -0.3(1) µm, standoff distance = 8.7(2) µm, magnetic dipole moment = 26(1)×10$^{-15}$ J/T, azimuthal orientation = 82(1)°, and polar orientation = 112(1)°. **i**, Decomposed ODMR spectra of an arbitrary pixel of the low-$\Delta_B$ measurement, showing estimated $c_1(f)$ (blue), $c_2(f)$ (orange), $c_3(f)$ (yellow), and $c_4(f)$ (purple), offset for clarity.

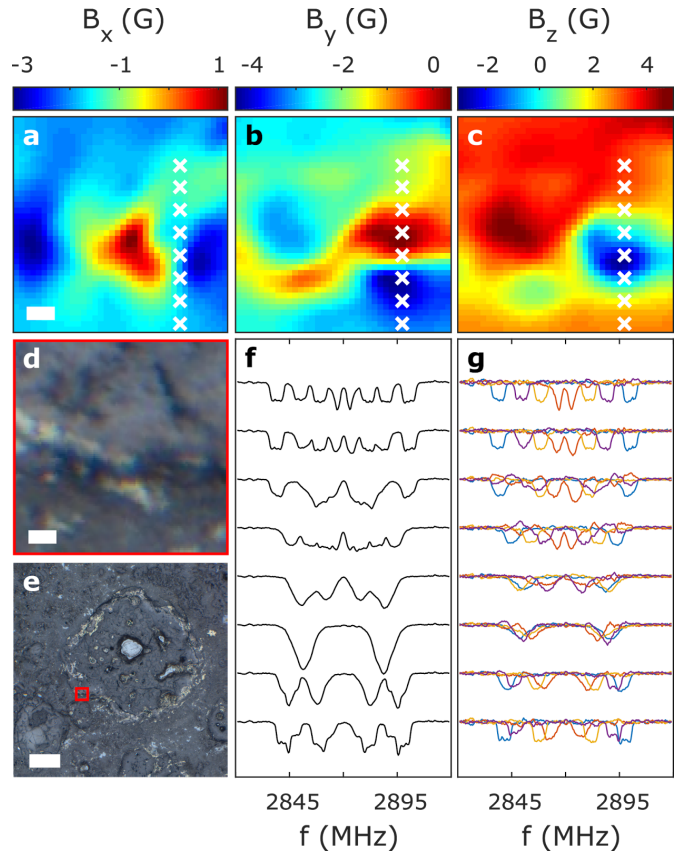

**Figure 4 | Meteorite magnetic imaging with Fourier optical decomposition. a-c**, Images of *x*, *y*, and *z* components of the magnetic field due to a subregion of the Allende meteorite sample, as determined using Fourier optical decomposition to resolve ambiguities from overlapping ODMR peaks. $\Delta_B$ = 8.42 G. Scale bar: 25 μm. **d**, Reflection brightfield image of same region of the meteorite as in **a-c**. Scale bar: 25 μm. **e**, Reflection brightfield image showing larger field-of-view around the region imaged in **a-d**, indicated by red square. Scale bar: 500 μm. **f**, Example ODMR spectra without Fourier decomposition in the pixels marked with "x"s in **a-c**. The eight resonances collapse into two or four features in some pixels, resulting in ambiguity. **g**, Fourier optical decomposition applied to the same ODMR spectra as in **f**, showing individual contributions due to only $c_1(f)$ (blue), $c_2(f)$ (orange), $c_3(f)$ (yellow), and $c_4(f)$ (purple).

Hyperfine features in **f** and **g** are blurred due to a combination of strong spatial gradients of the sample and a boxcar filter applied to improve SNR (see Methods). Spectra within the same panel in **f** and **g** are offset for clarity.

**Methods**

*Optical Simulation*

We simulated NV emission by adapting existing code[1] based on earlier works[2-6] modeling electric dipole emission near interfaces and collected with a high-NA objective. For the present work, we modeled an NV of a given orientation as two mutually incoherent radiating electric dipoles oriented perpendicular to the NV axis as sketched in Fig. 1c. We approximated monochromatic emission of wavelength 700 nm, near the peak of the NV$^-$ emission spectrum at room temperature[7]. Briefly, we computed the complex-valued electric field at the Fourier plane of the microscope by decomposing into constituent plane waves, each carrying a complex amplitude according to the emitter's orientation and defocus, as well as the appropriate Fresnel coefficients incurred as a result of transmission through the diamond-immersion medium interface. When modeling the detection of NVs located at the far surface of the diamond we also included Fresnel coefficients describing reflection from the far diamond-air interface. We performed simulations for both an NA 1.49/oil and NA 0.75/air objective, as both were used in our experiments. To approximate the two diamond samples used in our experiments, we simulated an NV ensemble containing equal populations of each of the four orientations, distributed uniformly throughout a surface layer with depth of either $d = 3.8$ μm or 0.9 μm (corresponding to samples D1 and D2 described below, respectively), with NV depth sampled every 10-100 nm. To match experimental conditions, the high-NA simulation of Fig. 1f places the NV layer at the (near) diamond-oil interface, while the low-NA simulation of Fig. S7 places the NV layer at the (far) diamond-air interface. The PSF shown in Fig. S2 was computed by taking the Fourier transform of the Fourier plane complex amplitude, then taking the square modulus. As was true in our imaging experiments, we assumed a magnification of 30x and a camera pixel size of 5.5 μm (183.3-nm pixels projected back to the object plane). The blocked and polarized PSF in Fig. S2 was computed by constraining support in the pupil plane to only x-polarized light in the right half of the Fourier disk, then taking the Fourier transform to propagate to the image plane.

*Experimental*

We used two diamond samples (made by Element Six), each grown by chemical vapor deposition. Sample D1 contains a surface NV layer of thickness 3.8 μm (as determined by secondary ion mass spectroscopy) and nitrogen concentration ~21 ppm. Sample D2 contains a surface NV layer of thickness 0.9 μm and nitrogen concentration ~7 ppm. Both diamond chips have dimensions 4 mm x 4 mm x 0.5 mm.

NVs were excited in a wide-field epi-illumination geometry, with 532-nm laser light coupled continuously into the back aperture of the objective via a dichroic mirror (Di02-R635, Semrock). The laser was linearly polarized along the axis $[1/\sqrt{2}, 1/\sqrt{2}, 0]^T$ (refer to coordinates in Fig. 1) such that each NV orientation was excited at equal rates. At the NV layer, laser beam peak intensity ranged ~7-360 W/cm². NV PL was collected with one of two objectives: 1) NA 1.49/oil

(CFI Apo TIRF 100x, Nikon), or 2) NA 0.75/air (CFI Plan Apo VC 20x, Nikon). Collected PL then was transmitted back through the aforementioned dichroic, and relayed to the camera by several mirrors and the train of lenses sketched in Fig. 1e. The tube lens TL ($f_{TL}$ = 300 mm) was placed a distance $f_{TL}$ from the back aperture of the objective, then the first 4$f$ lens L1 ($f_{L1}$ = 200 mm) was placed a distance $f_{TL} + f_{L1}$ from the tube lens. The Fourier plane (FP) is formed a distance $f_{L1}$ behind lens L1. To perform Fourier optical decomposition on the ODMR spectrum across an image, we placed an opaque knife edge mounted on a rotation mount at this FP. For imaging experiments, a second 4$f$ lens L2 ($f_{L2}$ = 200 mm) was then placed a distance $f_{L2}$ behind the FP, forming an image on the camera (acA2040-180km, Basler) a distance $f_{L2}$ behind L2. To image the Fourier plane directly as in Fig. 1g, we removed L2 and placed a Bertrand lens BL ($f_{BL}$ = 75 mm) a distance ~300 mm from the FP. This distance was determined first roughly using the thin lens equation, then fine-tuned to focus the FP at a common objective height as for the image. The linear polarizer was placed just before the BL, though it's precise placement is not important since the optical train behind the objective is well within the paraxial regime. A band-pass filter (Brightline Fluorescence Filter 726/128, Semrock) was placed on the outer aperture of the camera.

For the direct measurements of the Fourier plane depicted in Figs. 1 and 2 we used the NA 1.49/oil objective. The diamond D1 was oriented such that the NV layer was on the side facing the objective, directly in contact with the immersion oil. For the imaging experiments depicted in Figs. 3 and 4, this geometry could not be used since the magnetic sample has to be placed in close proximity to the NV layer. Thus the diamond (D1 for magnetic bead measurements, D2 for

Allende study) was flipped over and the objective focused through the bulk diamond. Imaging through the high-index ($n$ = 2.417) diamond has a significant effect on the microscope's PSF (Fig. S2). Because of its short working distance, the NA 1.49 objective could not be used to image through the diamond (thickness ~0.5 mm), and so imaging experiments were done instead with the NA 0.75/air objective. Future implementations can be done with a thinned diamond chip such that a high-NA objective can still be used, as the sensitivity of the ODMR decomposition improves with increasing NA (Fig. S4).

The external magnetic bias field was applied with a neodymium magnet mounted above the diamond. Microwaves were supplied with a TPI-1001-B synthesizer (Trinity Power, Inc.) and amplified with a ZHL-16W-43+ amplifier (Mini-Circuits), outputting ~44 dBm microwave power. Microwaves were delivered to the diamond via a copper wire loop oriented such that the microwave magnetic field had roughly equal projection on all four NV axes. The camera (operating at 200 fps) and microwave synthesizer were controlled with custom LabView and MATLAB software. Microwave power was toggled on/off in alternating images to help mitigate noticeable intensity drift from the laser. In each experiment we sampled 250 frequencies in random order, averaging for 50-250 microwave modulation periods. While the microwave lock-in improves SNR at a given frequency sample, the intensity drift can still limit the measurement as frequency switching was relatively slow: a fluctuation in laser power between two frequency samples causes fluctuations in relative contrast between the two samples. Our relatively inexpensive laser also exhibited random telegraph noise at times, and so some measurements were averaged up to 5 times to help alleviate this effect. The exact conditions in each measurement presented in the main figures were as follows: data in Figs. 1 and 2 were averaged for 100 microwave modulation periods per frequency sample, averaged once overall; unpolarized/unblocked data in Fig. 3 were averaged for 50 modulation periods per frequency sample, 5 times overall, while polarized/blocked data was averaged for 125 modulation periods per frequency sample, 4 times overall; unpolarized/unblocked data in Fig. 4 was averaged for 100 modulation periods per frequency sample, averaged once overall, and polarized/blocked data was averaged for 250 modulation periods per frequency sample, averaged once overall.

Before each measurement, diamond chips were cleaned by sonicating for 30 min in acetone, then 30 min in isopropyl alcohol. Ferromagnetic 2-μm diameter bead (Spherotech) samples were prepared by diluting 1/100 from stock, sonicating for 30 min, then pipetting an aliquot onto the NV layer surface of the diamond chip and leaving to dry on top of a permanent magnet in order to preferentially orient the beads at the surface.

Paleomagnetism measurements were done by placing the rock surface in contact with the NV layer surface of the diamond. Before magnetic imaging, we applied a 2000-G isothermal remanent magnetization (IRM) field to the Allende sample. This step allowed us to more easily identify magnetic sources for this proof-of-principle, and to simulate a highly magnetized meteorite sample in order to illustrate the dynamic range-extending capability of Fourier decomposition imaging over conventional vector imaging.

***Image Analysis***

For both Fourier- and real-space measurements, ODMR image data were stored as three-dimensional (two spatial and one microwave frequency) arrays to be analyzed with custom MATLAB software. To analyze Fourier-space images such as those in Figs. 1 and S7, each 237x237 pixel image was first smoothed with a Gaussian filter ($\sigma = 5$ pixels). A slight 4° rotation of the images due to subtle misalignments in the reflection axes of our mirrors was corrected in post-processing for analysis of the Fourier plane. We computed the elements of the transformation matrix **W** from these images, which by symmetry we expect to be of the form:

$$\mathbf{W} = \begin{pmatrix} w_1 & w_2 & w_3 & w_2 \\ w_2 & w_1 & w_2 & w_3 \\ w_3 & w_2 & w_1 & w_2 \\ w_2 & w_3 & w_2 & w_1 \end{pmatrix}. \tag{4}$$

In practice we did not enforce this symmetry, and instead used the experimentally measured matrix elements. For sample D1 measured with the 1.49/oil objective the experimentally measured matrix was:

$$\mathbf{W} = \begin{pmatrix} 0.222 & 0.355 & 0.058 & 0.358 \\ 0.351 & 0.213 & 0.354 & 0.060 \\ 0.062 & 0.371 & 0.222 & 0.359 \\ 0.365 & 0.060 & 0.367 & 0.222 \end{pmatrix}. \tag{5}$$

The matrix entries are similar to the simulated values quoted in main text. Simulations dictate that we should expect a different **W** when detecting NVs through the diamond using the 0.75/air objective. In this case we found simulated values of $w_1 = 0.175$, $w_2 = 0.370$, and $w_3 = 0.085$. Again in practice we used the experimentally measured transformation matrix, now given by:

$$\mathbf{W} = \begin{pmatrix} 0.178 & 0.366 & 0.090 & 0.374 \\ 0.363 & 0.172 & 0.364 & 0.085 \\ 0.094 & 0.368 & 0.176 & 0.375 \\ 0.366 & 0.094 & 0.370 & 0.166 \end{pmatrix}. \tag{6}$$

The above was used for lower-NA measurements of sample D1. The values changed very slightly for the thinner NV layer of sample D2.

For real-space measurements, the aforementioned 4° rotation was compensated by a commensurate rotation of the polarizer axis and knife edge away from horizontal/vertical. Rotating the polarizer and beam block between each $\tilde{c}_i(x_k, y_k, f)$ measurement caused small but measurable relative shifts of the images. To co-register the real-space images with one another we computed their cross-correlations then shifted to compensate the offset between the peaks of the correlation functions before further analysis. Blocking half of the pupil results in an elongated PSF (Fig. S2), which in turn means that the image of a point source recorded with the left/right half of the pupil blocked will not completely overlap with an image of the same point source recorded with the top/bottom of the pupil blocked. To compensate for this we performed Lucy-Richardson deconvolution on each slice of the ODMR image using the simulated PSF (again rotated by 4°) (Fig. S2) and the MATLAB function `deconvlucy`. While this deconvolution step appeared to improve magnetic bead images as determined by visual comparison to high-field images, it did not noticeably affect images of the rock's magnetic features, and so was not included in the analysis of the data presented in Fig. 4. The difference is likely explained by the fact that the magnetic features due to the rock were higher in magnitude and spatially broader than those of the beads. The deconvolution step is likely to be more important for small signals and spatial resolutions approaching the diffraction limit. In analyzing the unpolarized/unblocked data presented in Fig. 3b-d we included a deconvolution step using the appropriate unblocked simulated PSF (Fig. S2) for the sake of fair comparison.

We found that data taken with the Fourier decomposition method was somewhat sensitive to the objective's focal position. As shown in Fig. S2, the focal plane is ill-defined when imaging through the bulk diamond due to the appearance of sidelobes along the optical axis. The central spot of the simulated lateral PSF is in fact narrower when the objective is positioned at the second-brightest peak along $z$. Experimentally we also noted something resembling multiple foci, and seemed to find best results when positioned at the second-deepest such focal point. For deconvolution we used the PSF corresponding to the simulated second focal position (Fig. S2).

Since the features imaged in our studies did not necessitate such fine pixelation to resolve, we low-pass filtered the data by applying a Gaussian blur ($\sigma = 20$ pixels) and then binning (10x10 for magnetic bead imaging, 25x25 for rock imaging) the image at each frequency slice. The strong, localized magnetic features of the Allende section imbued steep magnetic field gradients on the NVs, causing significant broadening of the resonances. These broadened spectra were smoothed via a boxcar average of length 5 applied along the frequency axis before fitting. No such frequency boxcar was applied to the magnetic bead data.

The spectrum in each pixel was fit with specified lineshapes using least-squares fitting. For each NV orientation we fit each resonance lineshape as the sum of three Lorenztian functions separated by 2.16 MHz to account for $^{14}$N hyperfine splitting. The width, height, and central frequency of the Lorentzians were free parameters of the fit. Thus for each NV orientation this yields a total of either 4 (1 height, 1 width, and 2 positions) or 6 (2 heights, 2 widths, and 2 positions) free parameters—the difference between the two cases was insignificant. Once the positions of each of 2x4 = 8 resonances were extracted, they were fed to a least-squares fit of the Hamiltonian with parameters $B_x$, $B_y$, $B_z$, $M_{1z}$, $M_{2z}$, $M_{3z}$, and $M_{4z}$ (see below). An additional Gaussian blur was then applied to the resulting magnetic images ($\sigma = 0.5$ pixels for magnetic bead imaging, $\sigma = 1$ pixel for rock imaging).

The slowly-varying applied bias magnetic field was removed from images of magnetic beads by fitting the entire image (FOV ~ 150 μm x 150 μm) to a 4th order polynomial and subtracting the offset. The images of the magnetic bead shown in Fig. 3 are only a small subset of the mostly empty images recorded around it. A different approach to background subtraction was used for the rock sample since slowly varying recorded magnetic fields may be the result of real sources buried deeper within the rock. In this case we measured the background magnetic image due solely to the bias field by removing the rock slice from the diamond, then subtracted this resulting map from the rock magnetic images.

Finally, magnetic bead images were fit with least-squares to a magnetic dipole source image with 6 free parameters: $x$ position, $y$ position, standoff distance, magnetic dipole moment, azimuthal orientation, and polar orientation. The rock image shown in Fig. 4 was fit to two such dipole sources, plus a linearly varying background (Fig. S6).

*Hamiltonian Model*

The relevant spin Hamiltonian (in frequency units) for the NV oriented parallel to $\hat{\mathbf{u}}_1$ (Fig. 1a) is:

$$\begin{aligned}\mathcal{H}_1 &= (D + M_{z_1})\left(S_{z_1}^2 - \tfrac{2}{3}\right) + \gamma \mathbf{S_1} \cdot \mathbf{B} - M_{x_1}\left(S_{x_1}^2 - S_{y_1}^2\right) \\ &\quad + M_{y_1}\left(S_{x_1}S_{y_1} + S_{y_1}S_{x_1}\right) + \mathbf{S_1} \cdot \mathbf{A} \cdot \mathbf{I}\end{aligned} \quad (7)$$

In Eq. 7 $D = 2.87$ GHz is the zero-field splitting, $\mathbf{S_1}$ is the electronic spin operator, $\gamma = 2.8$ MHz/G is the electronic gyromagnetic ratio, $\mathbf{B}$ is the magnetic field, terms including components of $\mathbf{M_1}$ account for the spin-stress interaction[8, 9] (see below), $\mathbf{I}$ is the $^{14}$N nuclear spin operator, and $\mathbf{A}$ is the associated hyperfine tensor. The coordinate system $\{x_1, y_1, z_1\}$ is defined such that $z_1$ points along $\hat{\mathbf{u}}_1$, and $x_1$ coincides with one of the mirror planes of an NV with this orientation. Analogous Hamiltonians are defined for the other three NV orientations, with $\{x_j, y_j, z_j\} \forall j \in \{1, 2, 3, 4\}$ related to one another and to the global coordinates $\{x, y, z\}$ (Fig. 1) via the appropriate rotation matrices.

The components of $\mathbf{M_1}$ are related to the stress tensor $\bar{\bar{\sigma}}$ as described in references[8, 9], and to those of $\mathbf{M_j}$ for $j \neq 1$ via rotation matrix transformations of $\bar{\bar{\sigma}}$. Since $\bar{\bar{\sigma}}$ has 6 free parameters, the four equations of the form in Eq. 7 imply 3 magnetic + 6 stress = 9 total free parameters. The 8 resonance positions we fit in each ODMR spectrum means that the problem is under-determined without additional assumptions. To reduce the number of free parameters to 3 magnetic + 4 stress = 7 total, we neglect terms proportional to $M_{x_j}$ and $M_{y_j}$ yielding the simplified Hamiltonian:

$$\mathcal{H}_\mathbf{j} = \left(D + M_{z_j}\right)\left(S_{z_j}^2 - \tfrac{2}{3}\right) + \gamma \mathbf{S_j} \cdot \mathbf{B} + \mathbf{S_j} \cdot \mathbf{A} \cdot \mathbf{I} \tag{8}$$

for each *j*. This approximation is justified by considering a magnetic field oriented along $\hat{u}_j$ and treating $\mathbf{M_j}$ perturbatively. For the 0→ +1 transition, to second order:

$$f_{0\to +1} = D + \gamma B_{z_j} + M_{z_j} + \frac{M_{x_j}^2 + M_{y_j}^2}{2\gamma B_{z_j}} \tag{9}$$

For our work we expect only modest amounts of stress due to lattice imperfections, with $M_{x_j} \approx M_{y_j} \approx M_{z_j} \approx 0.1$ MHz. In this case, again considering the fact that the minimum Zeeman splitting due to the applied field was ~4 MHz in our studies, the third term on the RHS in Eq. 9 contributes a correction of ~0.1 MHz, while the fourth term would only contribute a correction of ~2.5 kHz. A future application at $B^{(\text{bias})} < \sim 1$ G may necessitate measuring each $\{M_{x_j}, M_{y_j}, M_{z_j}\}$ associated with the diamond region first without a magnetic sample of interest.

*Methods References*

1. Backlund, M. P. *et al*. Removing orientation-induced localization biases in single-molecule microscopy using a broadband metasurface mask. *Nature Photonics* **10**, 459-462 (2016).

2. Backer, A. S. & Moerner, W. E. Extending single-molecule microscopy using optical Fourier processing. *The Journal of Physical Chemistry B* **118**, 8313-8329 (2014).

3. Axelrod, D. Fluorescence excitation and imaging of single molecules near dielectric‑coated and bare surfaces: a theoretical study. *J. Microsc.* **247**, 147-160 (2012).

4. Richards, B. & Wolf, E. *Electromagnetic diffraction in optical systems. II. Structure of the*

# Supplementary Information

**Fourier optical processing enables new capabilities in diamond magnetic imaging**

Mikael P. Backlund[1,2], Pauli Kehayias[1,2], and Ronald L. Walsworth[1,2]

1. Harvard-Smithsonian Center for Astrophysics, Cambridge, Massachusetts 02138, USA
2. Department of Physics, Harvard University, Cambridge, Massachusetts 02138, USA

**Sensitivity Comparison**

The utility of our Fourier optical decomposition method arises when there is ambiguity in distinguishing the contributions from each NV orientation class to the overall ODMR spectrum. There is no advantage in signal-to-noise relative to the conventional approach to NV ensemble vector magnetometry if a high enough bias magnetic field can be applied such that the ODMR peaks are resolved. In fact, in this case Fourier decomposition would lead to diminished SNR that depends on the elements of the partition matrix $\mathbf{W}$. To illustrate this, suppose that a strong bias field is applied such that the resonances of each of the four NV orientations are well-separated. Let's consider the SNR at a microwave frequency that is on resonance with the NVs pointed along $\hat{\mathbf{u}}_1$, and off resonance for the other three classes. Thus the expected values for the signal due to each orientation are $\mathrm{E}(C_1) = c_1$ and $\mathrm{E}(C_{j \neq 1}) = 0$. Suppose we measure each component of the vector $\tilde{\mathbf{C}}$ by applying the appropriate Fourier filter as described in the main text, and that we seek to estimate the value of $C_1$ by the following transformation:

$$\hat{C}_1 = [1, 0, 0, 0] \cdot \mathbf{W}^{-1} \tilde{\mathbf{C}} \tag{S1}$$

We are interested in the measurement uncertainty of $\hat{C}_1$. From Eq. S1 we have:

$$\mathrm{E}(\hat{C}_1) = [1,0,0,0] \cdot \mathbf{W}^{-1}\mathrm{E}(\tilde{\mathbf{C}}) = [1,0,0,0] \cdot \mathbf{W}^{-1} \begin{pmatrix} w_1 \\ w_2 \\ w_3 \\ w_2 \end{pmatrix} c_1 = c_1 \quad (S2)$$

as it should be. To determine $\mathrm{Var}(\hat{C}_1)$ we define the covariance matrix $\tilde{\Sigma}$ of $\tilde{\mathbf{C}}$. Let $\tilde{\Sigma} = \sigma^2 \mathbf{I}$, where $\mathbf{I} \in \mathbb{R}^{4x4}$ is the identity matrix. The matrix $\tilde{\Sigma}$ is diagonal because each measurement $\tilde{C}_i$ is independent. We assume $\mathrm{Var}(\tilde{C}_i)$ is the same $\forall i$ (and is equal to $\sigma^2$). This assumption is approximately valid for Poisson $\tilde{C}_i$ in the limit of low (~1%) ODMR contrast. In the presence of other noise sources (e.g., readout electronics), the approximation is likely even better. The assumption simplifies the comparison below. We can now write down the variance of our estimator:

$$\mathrm{Var}(\hat{C}_1) = [1,0,0,0] \cdot \mathbf{W}^{-1}\tilde{\Sigma}\mathbf{W}^{-1} \cdot \begin{bmatrix} 1 \\ 0 \\ 0 \\ 0 \end{bmatrix} = \sigma^2 \{\mathbf{W}^{-2}\}_{11} \quad (S3)$$

The notation $\{\mathbf{W}^{-2}\}_{11}$ here means the (1,1) element of the matrix $(\mathbf{W}^{-1})^2$. Using the definition of $\mathbf{W}$ given in the text, and noting that $w_2 = (1 - w_1 - w_3)/2$, this gives:

$$\mathrm{Var}(\hat{C}_1) = \frac{\sigma^2}{4}\left(1 + \frac{2}{(w_1-w_3)^2} + \frac{2}{(2(w_1+w_3)-1)^2}\right) \quad (S4)$$

We define the sum and difference parameters $w_- = w_1 - w_3$ and $w_+ = w_1 + w_3$. Then Eq. S4 can be rewritten:

$$\mathrm{Var}(\hat{C}_1) = \frac{\sigma^2}{4}\left(1 + \frac{2}{w_-^2} + \frac{2}{(2w_+-1)^2}\right) \quad (S5)$$

Let us compare this result to the SNR expected for a conventional NV ensemble vector magnetometry measurement, in which we measure the variable $C_{\text{tot}} = \sum \tilde{C}_i$ with the following statistics: $\text{E}(C_{\text{tot}}) = c_1$ and $\text{Var}(C_{\text{tot}}) = 4\sigma^2$. Thus the relative SNR is:

$$\sqrt{\frac{\text{Var}(C_{\text{tot}})}{\text{Var}(\hat{C}_1)}} = 4\left(1 + \frac{2}{w_-^2} + \frac{2}{(2w_+ - 1)^2}\right)^{-1/2} \qquad (S6)$$

As implemented in the current experiment, the realization of $C_{\text{tot}}$ takes ¼ the amount of time as the realization of $\hat{C}_1$, and so the RHS of Eq. S6 should have an addition factor of ½ in this case. However, a future implementation of Fourier decomposition can be parallelized with appropriate beamsplitters and multiple detectors, and so Eq. S6 describes the ultimate limit of such a method. Furthermore, an additional time factor >1 can be added to the RHS of Eq. S6 in cases where a low bias field allows for a narrower sweep in microwave frequency, assuming fixed frequency increments. In our experiment this was not the case, however, as we instead fixed the number of frequency points to 250 for both high and low bias field measurements such that the low field case had finer frequency sampling.

Figure S4 shows the relative SNR defined in Eq. S6 for the physically relevant range $0 < w_- < w_+ < ⅓$. The values of $w_-$ and $w_+$, of course, depend on the NA and immersion medium. Results for the two objectives used in this study (1.49/oil and 0.75/air) are highlighted in Fig. S4, yielding a reduction in SNR by factors of 0.47 and 0.25, respectively. In general, better SNR is attained when $w_-$ and $w_+$ are both maximized, which coincides with large NA. In the limit $w_- = 0$ we have $w_1 = w_3$, and thus pairs of NV orientations can no longer be distinguished by this method and the matrix **W** becomes singular.

**Figures**

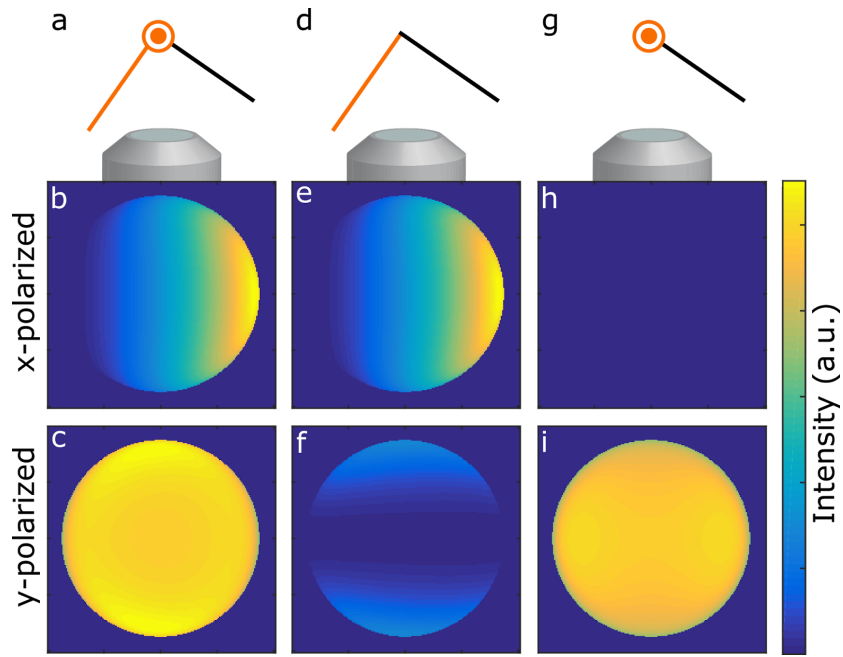

**Figure S1 | Intuitive explanation of Fourier plane pattern. a,** Diagram depicting an NV oriented along $\hat{u}_1$ (black line) and it's two transition electric dipole moments (orange) above the microscope objective. **b,** Intensity of the *x*-polarized light emitted from both transition dipoles depicted in **a**. **c,** Intensity of the *y*-polarized light emitted from both transition dipoles depicted in **a**. The patterns in **b, c** differ very slightly from those shown in Fig. 1f of the main text as Fig. 1f illustrates contrast rather than intensity, and so is divided by the total intensity pattern (which has a slight spatial dependence) in each polarization channel. **d,** Sketch depicting only one of the transition dipoles. Since light is emitted with polarization mostly parallel to the dipole, the *x*-polarized channel will have most of the light due to this dipole. Since the dipole emission pattern has a null in the direction parallel to the dipole, and since this dipole is tilted out of the sample plane, the distribution of light collected in the Fourier plane will vary from left to right. **e,**

Intensity of the *x*-polarized light emitted only from the dipole depicted in **d**. **f,** Intensity of the *y*-polarized light emitted only from the dipole depicted in **d**. **g,** Sketch depicting only the other transition dipole. Since light is emitted with polarization mostly parallel to the dipole, the *y*-polarized channel will have most of the light due to this dipole. Since this dipole lies parallel to the sample plane, the distribution of light collected in the Fourier plane will be nearly uniform. **h,** Intensity of the *x*-polarized light emitted only from the dipole depicted in **g**. **i,** Intensity of the *y*-polarized light emitted only from the dipole depicted in **g**. While the colorbar units are arbitrary, the scale is consistent across all panels.

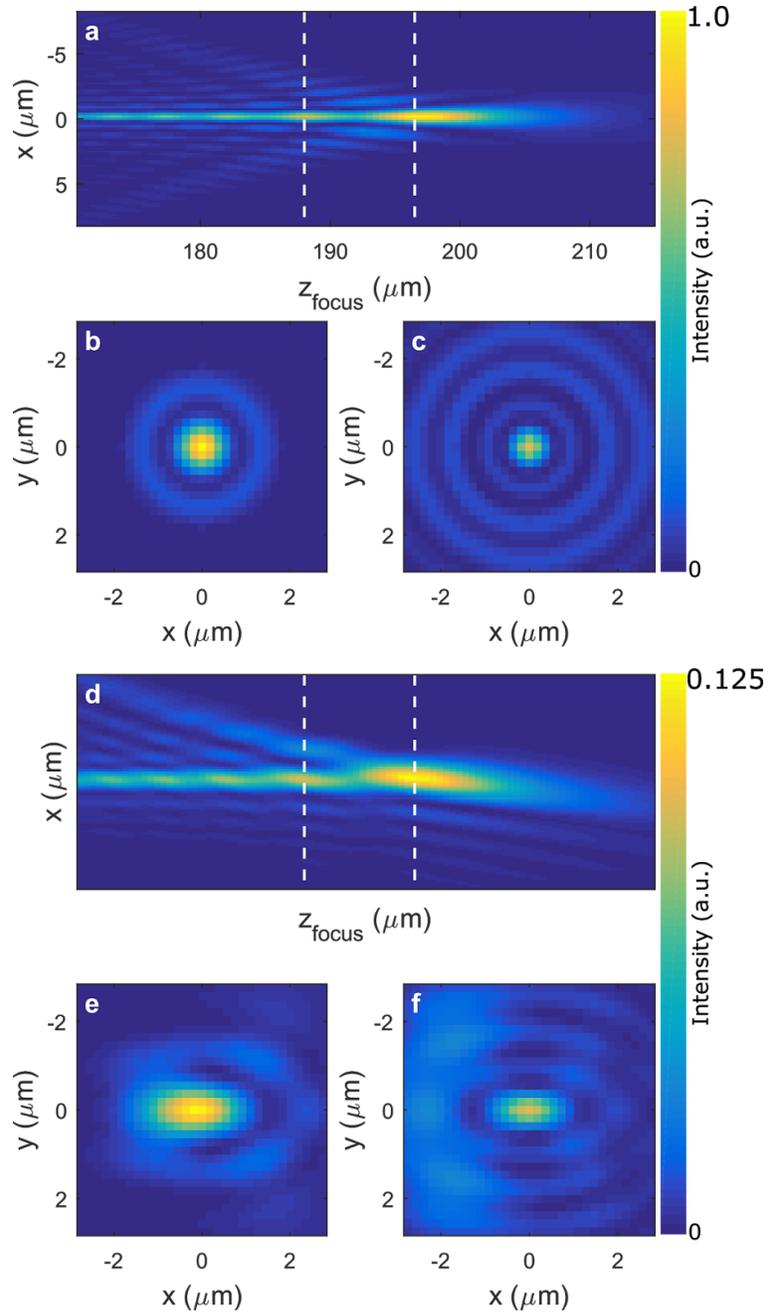

**Figure S2 | Simulated PSF.** Assumes an NA 0.75/air objective imaging an NV layer of thickness 3.8 μm on the far side of a 0.5-mm thick diamond chip. **a**, *xz* slice of 3D PSF. $z_{focus}$ is the distance of the nominal focal plane of the objective above the diamond-immersion oil interface. Despite the fact that the NV layer is ~500 μm above the diamond-oil interface, the

layer appears in focus around $z_{focus}$ = 200 μm due to the severe focal shift resulting from the large index of refraction mismatch. Intense side lobes on the near side of the PSF make the "focal plane" somewhat ill-defined. The white dashed lines mark the primary focal plane at which the 2D PSF is brightest (right), as well as a secondary focal plane at which the central lobe of the 2D PSF is actually narrower (left). **b**, 2D slice of unblocked PSF at primary focal plane. **c**, 2D slice of unblocked PSF at secondary focal plane. **d**, *xz* slice of the PSF resulting from blocking the left half of the Fourier plane. Note the different color scale. **e**, 2D slice of blocked PSF at primary focal plane. **f**, 2D slice of blocked PSF at secondary focal plane.

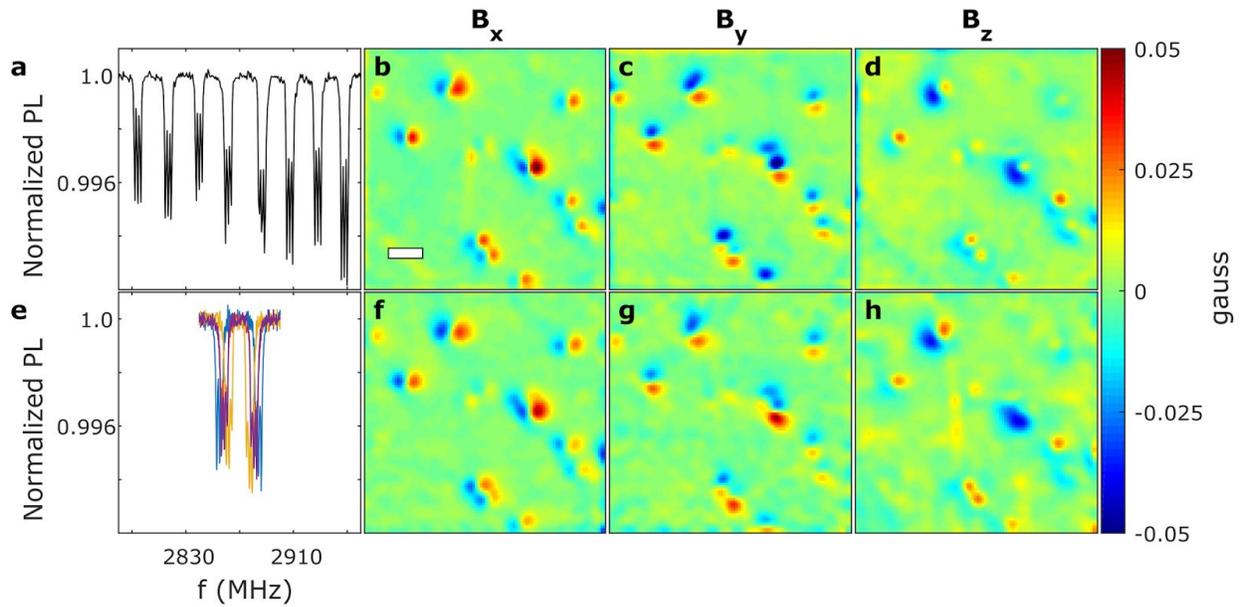

**Figure S3 | Additional magnetic bead images.** An aliquot of magnetic bead mixture was dropcast on the diamond surface, then the diamond was allowed to dry on top of a permanent magnet such that the beads tended to orient along the same direction. **a**, Spatially-averaged ODMR taken at $\Delta_B$ = 22.96 G such that all eight resonances are resolved. **b-d**, Images of

magnetic field components as measured at $\Delta_B = 22.96$ G without Fourier optical decomposition. **e**, Spatially-averaged ODMR taken at $\Delta_B = 2.47$ G, after Fourier optical decomposition. Colors correspond to different NV orientations as indicated in main text. **f-h**, Images of magnetic field components of same field-of-view, as measured at $\Delta_B = 2.47$ G after Fourier optical decomposition. Subtle differences between **b-d** and **f-h** in the magnetic images of certain beads may be in part due to physical reorientation of the beads, as the orientation and magnitude of the bias field is different in the two sets of images. Image scale bar: 25 µm.

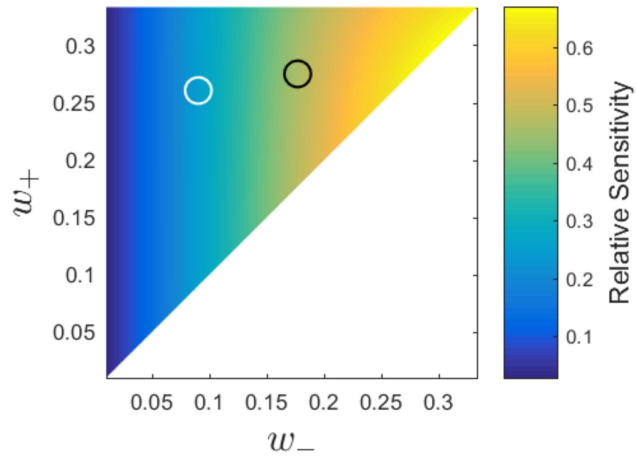

**Figure S4 | Relative sensitivity of ODMR techniques.** Sensitivity of Fourier optical decomposition method relative to conventional NV ensemble vector magnetometry, as a function

of $w_-$ and $w_+$, according to Eq. S6. Black circle highlights the relevant value for NA 1.49/oil, and white circle for NA 0.75/air.

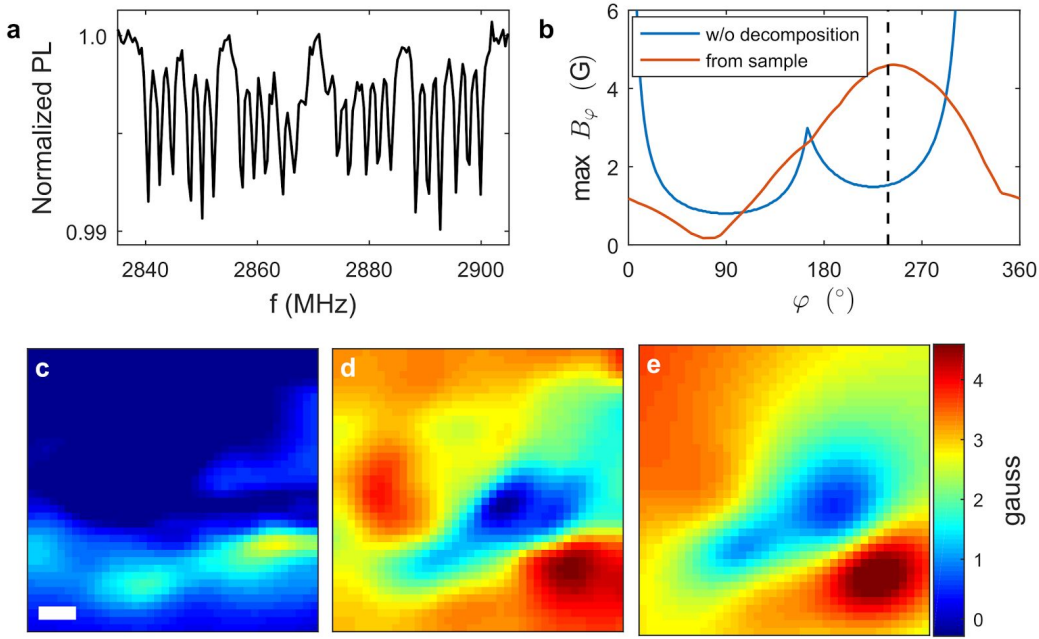

**Figure S5 | Dynamic range quantification.** Another look at data from the measurement of the Allende slice, here highlighting the extension of dynamic range provided by our method. **a,** ODMR spectrum obtained from diamond before placing the sample. The chosen bias field has $\Delta_B$ = 8.42 G, just large enough to clearly resolve the resonances. **b,** A comparison of the dynamic range afforded by the chosen $\mathbf{B}^{(bias)}$ for a conventional magnetic imaging experiment without Fourier decomposition (blue), and the dynamic range required by the sample (orange). The vertical axis is the maximum magnitude of the magnetic field applied along the in-plane direction $\varphi = \arctan(y/x)$. The blue curve was calculated by computationally increasing $B$ at

each $\varphi$ until at least two resonances in the ODMR are separated by twice the hyperfine splitting, $2 \times 2.16 = 4.32$ MHz, at which point ambiguities between resonances would arise. This may be a conservative estimate of dynamic range since, as in the case of the Allende measurement, the resonances may become significantly broadened by large spatial gradients. The orange curve was computed from the field due to the sample as determined with Fourier decomposition. The dynamic range required by the sample at $\varphi = 239°$ (black dashed line) is more than three times the dynamic range of a conventional measurement at this bias field. **c,** nonsense image of $\hat{B}_{\varphi=239°} = \hat{B}_x \cos(239°) + \hat{B}_y \sin(239°)$ obtained by attempting to fit spectra without Fourier optical decomposition. **d,** Image of $\hat{B}_{\varphi=239°}$ obtained by fitting spectra after Fourier optical decomposition. **e,** Expected $B_{\varphi=239°}$ determined from fit dipolar fields described below in Fig. S6. Scale bar for images: 25 μm.

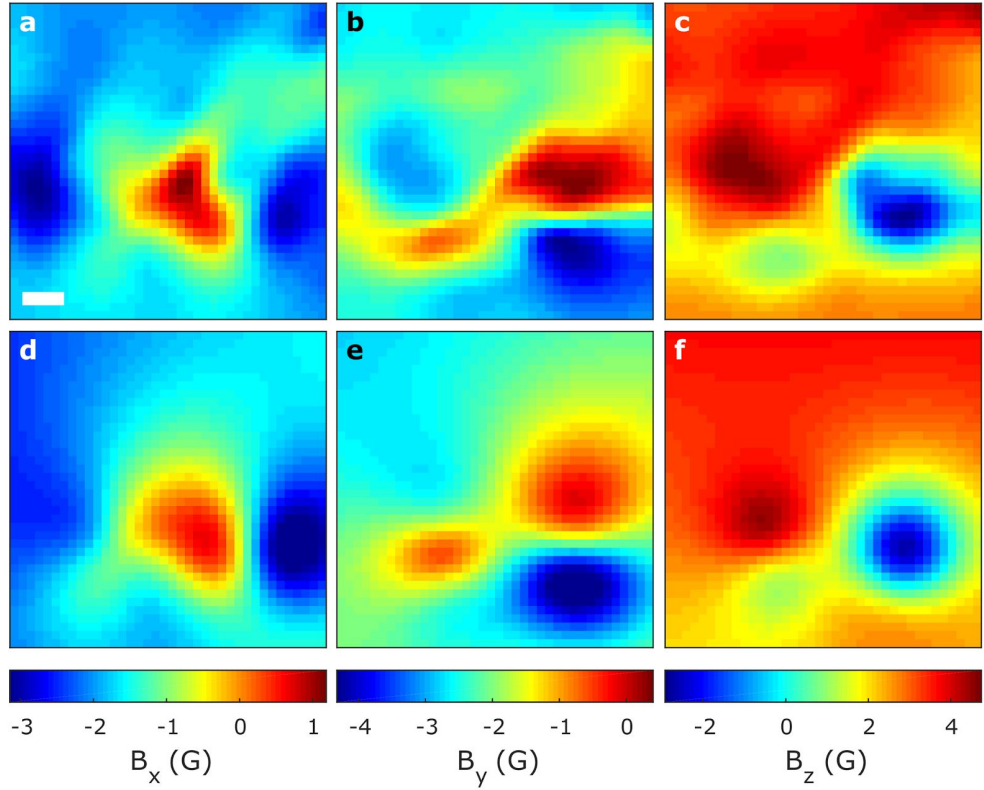

**Figure S6 | Least-squares fit to Allende magnetic images. a-c**, Same panels as Fig. 4a-c in main text, i.e., images of $B_x$, $B_y$, and $B_z$ due to region of Allende sample, as determined after Fourier optical decomposition of the ODMR data to resolve ambiguities. **d-f**, Least-squares fit of images in top row to a model consisting of two magnetic dipoles plus a linearly varying background. Fit parameters for each dipole were $x$ position, $y$ position, standoff distance, magnetic moment, azimuthal orientation, and polar orientation. The background in each component was modeled as a plane with 3 free parameters: an $x$ coefficient, a $y$ coefficient, and a constant offset. In all there are 21 free parameters in the fit. Notable results of fit: Dipole 1 has a standoff distance of 15.6(7) μm and magnetic moment $7.3(8) \times 10^{-12}$ J/T, while Dipole 2 has a standoff distance of 19.5(4) μm and magnetic moment $2.3(1) \times 10^{-11}$ J/T. Scale bar: 25 μm.

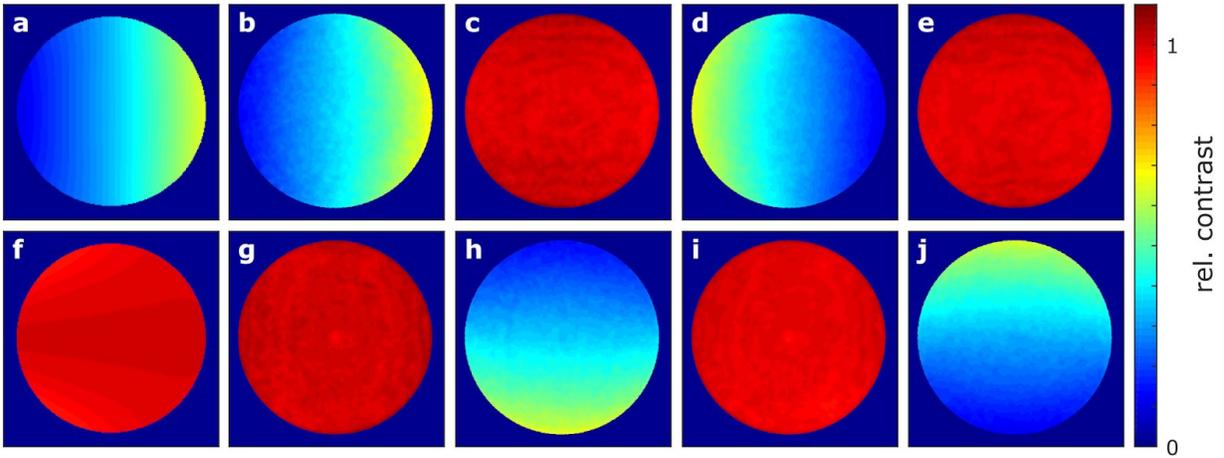

**Figure S7 | Simulated and experimental Fourier patterns at low NA.** Plots are analogous to Fig. 1f, g of the main text, except for detection is through the diamond chip using an NA 0.75/air objective. **a,** Simulated contrast distribution in *x*-polarized channel due to NV oriented along $\hat{\mathbf{u}}_1$. **b-e**, Experimentally measured Fourier patterns in *x*-polarized channel for NVs oriented along $\hat{\mathbf{u}}_1, \hat{\mathbf{u}}_2, \hat{\mathbf{u}}_3, \hat{\mathbf{u}}_4$, respectively. **f,** Simulated contrast distribution in *y*-polarized channel due to NVs oriented along $\hat{\mathbf{u}}_1$. **g-j**, Experimentally measured Fourier patterns in *y*-polarized channel for NVs oriented along $\hat{\mathbf{u}}_1, \hat{\mathbf{u}}_2, \hat{\mathbf{u}}_3, \hat{\mathbf{u}}_4$, respectively.